\documentclass[twocolumn,showpacs,preprintnumbers,amsmath,amssymb,subeqn,superscriptaddress]{revtex4}
\usepackage{graphicx}
\usepackage{dcolumn}
\usepackage{bm}
\begin{document}
\title{How to make a fragile network robust and vice versa}
\author{Andr\'e A. \surname{Moreira}}
\email{auto@fisica.ufc.br} \affiliation{Departamento de
F\'{\i}sica, Universidade Federal do Cear\'a, 60451-970 Fortaleza,
Cear\'a, Brazil}
\author{Jos\'e S. Andrade Jr.}
\author{Hans J. Herrmann}
\affiliation{Departamento de F\'{\i}sica, Universidade Federal do
Cear\'a, 60451-970 Fortaleza, Cear\'a, Brazil}
\affiliation{Computational Physics, IfB, ETH-H\"onggerberg,
Schafmattstrasse 6, 8093 Z\"urich, Switzerland}
\author{Joseph O. Indekeu}
\affiliation{Instituut voor Theoretische Fysica, Katholieke
Universiteit Leuven, B-3001 Leuven, Belgium}
\date{\today}
\begin{abstract}
We investigate topologically biased failure in scale-free networks
with degree distribution $P(k) \propto k^{-\gamma}$. The
probability $p$ that an edge remains intact is assumed to depend
on the degree $k$ of adjacent nodes $i$ and $j$ through
$p_{ij}\propto(k_{i}k_{j})^{-\alpha}$. By varying the exponent
$\alpha$, we interpolate between random ($\alpha=0$) and
systematic failure. For $\alpha >0 $ ($<0$) the most (least)
connected nodes are depreciated first. This topological bias
introduces a characteristic scale in $P(k)$ of the depreciated
network, marking a crossover between two distinct power laws. The
critical percolation threshold, at which global connectivity is
lost, depends both on $\gamma$ and on $\alpha$. As a consequence,
network robustness or fragility can be controlled through fine
tuning of the topological bias in the failure process.
\end{abstract}
\pacs{64.60.aq, 
    89.75.Hc  
         64.60.ah, 
} 

\keywords{} 
\maketitle 

Scale-free networks, with power-law degree distribution
$P(k) \propto k^{-\gamma}$, are remarkably resistant to
random failure~\cite{barab,havlin1}. This quality is
important when failure is to be avoided, as in the
air-transportation network. It has been speculated that,
also in nature, scale-free design evolves as a way to
achieve robustness~\cite{evol}. On the other hand,
robustness may be a problem when one tries to halt an
epidemic. The fundamental question we ask and answer in this
work is how one can delicately control whether a network is
fragile or robust.

Previous work has mostly concentrated on homogeneous networks, in
which all edges have the same chance to fail. However, by design
or evolution the most critical edges of the network may become
less prone to failure. Also, a targeted attack can disrupt the
network after only a small fraction of edges
fail~\cite{barab,havlin3}. This shows that in heterogeneous
networks the topology alone does not determine the susceptibility
of breakdown.

The critical properties of static phenomena and dynamical
processes are affected by the topology of the network of
interactions~\cite{dorog}. It was recently
shown~\cite{indekeu1,indekeu2} that, by accounting for a topology
dependence in the interaction strength between the nodes,
$J_{ij}\propto (k_ik_j)^{-\alpha}$, one obtains a critical
behavior that mimics the case of homogeneous interaction but with
a different degree distribution. The system with exponent $\gamma$
and topology-dependent interactions can be mapped to a homogeneous
one, $\alpha=0$, but with an effective exponent
$\gamma^{\prime}$, given by~\cite{indekeu1,indekeu2}
\begin{equation}
\gamma^{\prime}=\frac{\gamma-\alpha}{1-\alpha}. \label{eq.indek}
\end{equation}

We focus on failure in scale-free complex networks, mediated
by a dynamical process that depends on the network topology.
Disregarding the presence of correlations, any such
dependence has to be related only to the node degree $k$. We
have to choose between two possible approaches, namely
failure of the nodes or, as we implement here, {\em failure
of the edges}. We express the failure
probability for an edge between nodes $i$ and $j$ as
$q_{ij}=q_{ij}(k_i,k_j)$. We assume that the network
depreciation occurs through a probability of occupation of
an edge $p_{ij}=1-q_{ij}$, which depends on the degrees of
the vertices,
\begin{equation}
p_{ij} \propto w_{ij}=(k_ik_j)^{-\alpha}, \label{eq.pij}
\end{equation}
where $w_{ij}$ is the topology-dependent weight of the edge.
Equation~(\ref{eq.pij}) is in the same spirit as the
degree-dependent interaction proposed in~\cite{indekeu1,indekeu2}.
Since failure can be related to the purely geometrical model of
percolation, its understanding does not require interactions but
can be achieved directly in terms of topological properties.

A topology-dependent depreciation allows to interpolate
smoothly between random failure ($\alpha =0$) and
intentional attack of links between hubs ($\alpha > 0$), or
intentional depreciation of edges between the least
connected nodes ($\alpha < 0$). We shall see that $\alpha
\in [2-\gamma,1]$ defines the useful range of topological
bias in the context of scale-free networks with finite mean
degree ($\gamma > 2$). Degree-dependent failure was also
studied in~\cite{gallos} for the case of node removal, while
edge removal was investigated in~\cite{wu}.

For uncorrelated networks, homogeneous random failure
($\alpha = 0$) can be solved using a mean-field
approach~\cite{havlin1,newman1,newman2,havlin2}. Close to
the critical fraction of occupied edges $f_c$, the size of
the largest connected cluster grows as $(f-f_c)^\beta$,
where the critical exponent $\beta$ depends on
$\gamma$~\cite{havlin2}. For $2<\gamma\le{3}$ the critical
point vanishes, $f_c \to 0$. As a consequence, we may say
that networks with $\gamma\le{3}$ are {\it robust} while
networks with $\gamma>3$ are {\it fragile}.  All these
results, however, are only relevant for the case of random
failure.  Henceforth, we will call the regime $\alpha>0$
``centrally biased'' ({\bf{CB}}). The converse regime,
$\alpha<0$, will be termed ``peripherally biased''
({\bf{PB}}).

To build our scale-free networks, we use the configuration model
(CM)~\cite{newman2}. The parameters of this model are the exponent
$\gamma$, the number of nodes (or vertices) $N_v$, and the minimum
degree allowed $k_{min}$. Unless said otherwise, all the networks
studied in this work have $k_{min}=2$. Depending on these
parameters and on the particular realization, we obtain a different
number of edges $N_e$. In this model, the degrees of the nodes are
determined initially from the desired distribution and then
connections are assigned at random.

To study the depreciation process, we start from a total failure
scenario, i.e., with all edges being initially removed from the
network. We then gradually include the edges back, with
probability proportional to some weight $w_{ij}$ that we will
assume to follow Eq.~(\ref{eq.pij}). By stopping this process at
intermediate steps, we can obtain results for the percolation
problem as the fraction of occupied edges grows from zero to one.

\begin{figure}[t]
\includegraphics*[width=7.0cm]{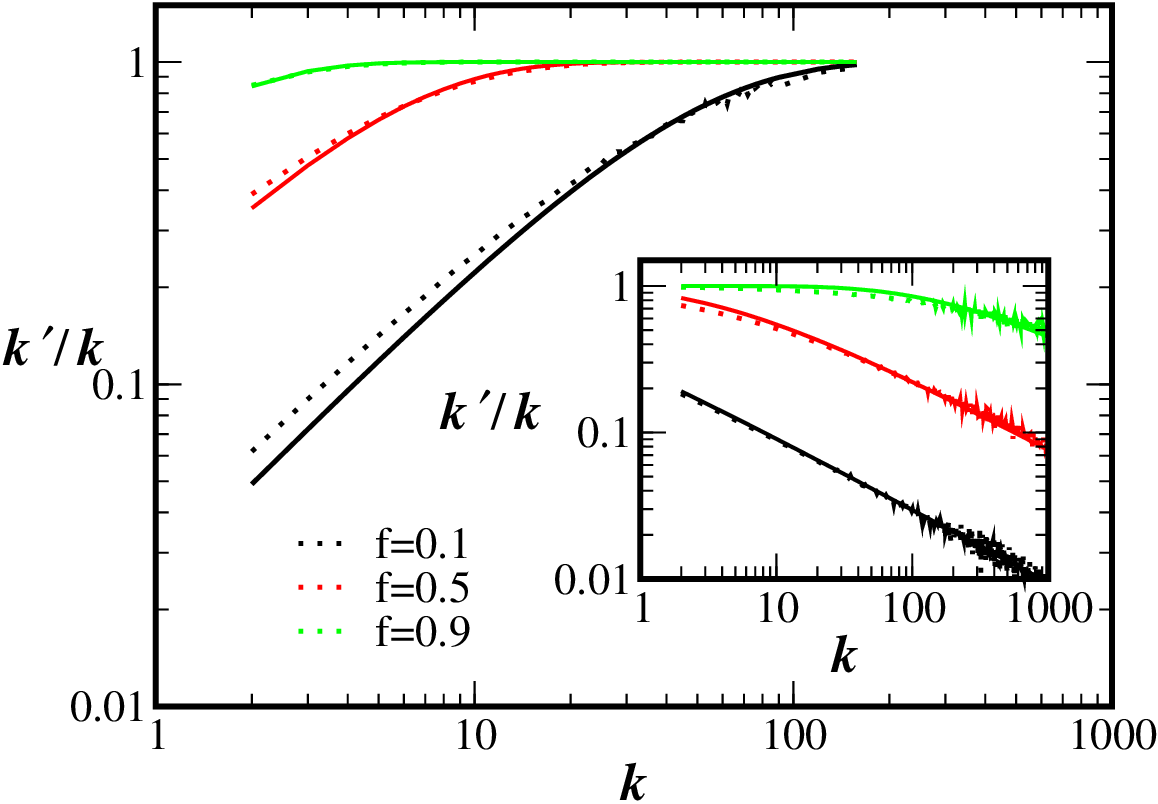}
\caption{Average degree reduction in the depreciated network as a
function of the original node degree. The main panel shows results
for networks with $\gamma=4$ submitted to PB with $\alpha=-1$. One
can see that the fraction $k^\prime/k$ grows as a power law with
degree $k$, saturating at $1$ at a scale that depends on the
fraction $f=n/N_e$ of edges in the depreciated network. The dotted
lines are the numerical results obtained from 10 network
realizations of size $N_v=10^5$. For each one of these networks,
the depreciation has been applied 10 times. The continuous lines
are the best fit to the data of Eq.~(\ref{eq.kprime}) with the
free parameter $C(f)=0.025, 0.22$ and $0.92$ for $f=0.1, 0.3$ and
$0.9$, respectively. The inset shows the same as in the main
panel, but for a network with $\gamma=2.5$ subjected to CB with
$\alpha=0.5$. In this case, the fraction $k^\prime/k$ decays with
$k$. For small scales one may find a saturation depending on the
fraction of edges in the depreciated network. The continuous lines
are the best fits to the data of Eq.~(\ref{eq.kprime}) with values
$C(f)=0.28, 2.5$ and $19$ for $f=0.1, 0.3$ and $0.9$,
respectively.} \label{f.degree}
\end{figure}

We now determine the probability $P_{ij}(f)$ that a particular
edge connecting nodes $i$ and $j$ is present in the network after
a fraction $f$ of the edges has already been included. This
probability can be identified as
$P_{ij}(f)=1-\prod_{t=1}^{n}{(1-w_{ij}/Z_t)}$, where $Z_t$ is the
mean (over the inclusion process) of the sum of weights of all
edges that have not yet been included in the network at step $t$,
and $n={f}N_e$ is the number of included edges. Assuming
$w_{ij}\ll{Z_t}$, we can write,
\begin{equation}
P_{ij}(f) \approx 1-e^{-D(f)(k_ik_j)^{-\alpha}},
\label{eq.kasteley}
\end{equation}
where the parameter $D(f)$ can be determined using
$\sum{P_{ij}=n}$. Using $w_{ij}$ as defined in Eq.~(\ref{eq.pij}),
the Kasteleyn-Fortuin construction~\cite{kastelen} allows us to
draw a parallel between the probability $P_{ij}$ and the
degree-dependent interaction previously proposed
in~\cite{indekeu1,indekeu2}.

We now define $\rho_k(f)$ as the mean probability that an edge
from a node with degree $k$ is present in the depreciated network.
We then have to average $P_{ij}$ over the nearest neighbors of a
node with degree $k$,
$\rho_k=\int_{k_{min}}^{\infty}{P_n(k_n)\left(1-\exp\left[-D(f)(k
k_n)^{-\alpha}\right]\right) dk_n}$.  For uncorrelated networks
the degree distribution of a neighbor is given by
$P_n(k_n)=P(k_n)k_n/\langle{k}\rangle$. Performing the integration
and examining the asymptotic behavior of the resulting incomplete
Gamma function, we find that $\rho_k$ is well approximated by
\begin{equation}
\rho_k \approx 1-e^{-C(f)k^{-\alpha}}, \label{eq.kprime}
\end{equation}
with
$C(f)=\frac{k_{min}^{-\alpha}(\gamma-2)}{\gamma-2+\alpha}D(f)$,
provided $\alpha\in\left[2-\gamma,1\right]$. The same range of
$\alpha$ is also featured in previous work on networks with
degree-dependent interactions~\cite{indekeu1,indekeu2}.
Equation~(\ref{eq.kprime}) is confirmed by the numerical results
shown in Fig.~\ref{f.degree}.

\begin{figure}[t]
\includegraphics*[width=7.0cm]{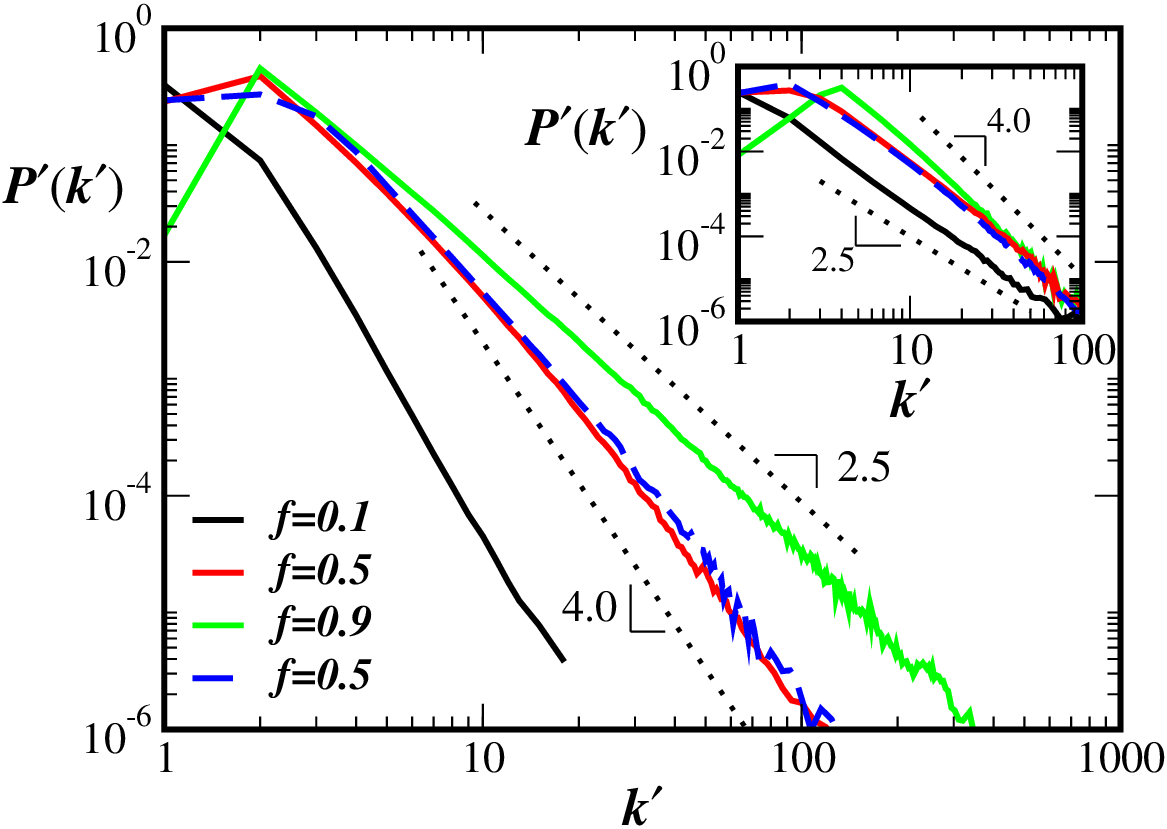}
\caption{Degree distribution of the depreciated network. This
result was obtained for 10 network realizations of size
$N_v=10^5$. For each network realization, the depreciation has
been applied 10 times. In the main panel we show the degree
distribution for networks with $\gamma=2.5$ submitted to CB
failure with $\alpha=0.5$. From Eq.~(\ref{eq.indek}) we expect the
value $\gamma^\prime=4.0$ for the depreciated network. As a guide
to the eye, the dotted lines indicate the power-law decays with
exponents 2.5 and 4. One can see that for small degree the
distribution initially decays with a slope very close to $\gamma$,
and then crosses over to a decay with a slope close to
$\gamma^\prime$ at a scale that depends on the fraction $f$ of
edges in the depreciated network. This shows that the
topology-dependent failure process introduces a characteristic
scale in the degree distribution of the originally scale-free
network. In the inset we show results for networks with $\gamma=4$
subjected to PB failure with $\alpha=-1$, which is equivalent to
an exponent $\gamma^\prime=2.5$. Contrary to the result of the
main panel, one sees a crossover from a slope $\gamma^\prime$ at
small degree to a slope $\gamma$ at large degree. The minimum
degree was set to $k_{min}=4$ in the results of the inset and
$k_{min}=2$ in the main panel. Surprisingly, for the case $f=0.5$
the degree distributions in the inset and the main panel are
remarkably similar. To illustrate this similarity we included the
results for $f=0.5$ from the main panel in the inset and vice
versa; these are the dashed (blue) lines. Although the two cases
start with distinct degree distributions at $f=1$, we obtain
similar distributions at a certain point of the depreciation
process.} \label{f.dist}
\end{figure}

Equation.~(\ref{eq.kprime}) can be used to determine the average
degree of a node after depreciation,
$k^\prime(k)=k~\rho_k$. From that we can obtain the degree
distribution of the depreciated network:
$P^\prime(k^\prime)={k^\prime}^\gamma$ for
$C(f)k^{-\alpha}\gg{1}$, and
$P^\prime(k^\prime)={k^\prime}^{\gamma^\prime}$ for
$C(f)k^{-\alpha}\ll{1}$, where $\gamma^{\prime}$ is given by
Eq.~(\ref{eq.indek}). We find that the degree distribution
after depreciation exhibits a crossover at a scale given by
$k_s\propto C(f)^{1/\alpha}$. As expected, the crossover is
not present in the random failure case, $\alpha=0$. However,
if the failure process is affected by the topological
properties of the network, as modeled by Eq.~(\ref{eq.pij}),
we have a characteristic scale $k_s$ that has not been
observed before. The presence of this crossover is supported
by the numerical results shown in Fig.~\ref{f.dist}.  It is
interesting to note that whether $\gamma$ or
$\gamma^\prime$~controls the decay at large degree depends
on the sign of $\alpha$. If $\alpha>0$ (CB), we have
$\gamma<\gamma^\prime$ and $\gamma^\prime$ controls the
decay at large degree, while for $\alpha<0$ (PB) the larger
exponent, $\gamma$, is the controlling one. Thus, the
largest of the two exponents $\gamma$ and $\gamma^\prime$
controls the asymptotic decay.  A robust network with
$\gamma\le{3}$ under CB and a fragile network with
$\gamma>3$ under PB may result in networks with similar
degree distributions after depreciation. The numerical
results shown in  Fig.~\ref{f.dist} correspond to the degree
distributions of networks under CB and PB failure.

Next we investigate the critical behavior associated with
percolation. A network is above the critical point when a
node connected to another node in the spanning cluster has
on average at least one other connection, thus assuring that
the cluster does not fragment. For an uncorrelated network,
this condition is equivalent to
$\langle{k^\prime}^2\rangle/\langle{k^\prime}\rangle>2$~\cite{havlin1}.
In order to determine the critical fraction $f_c$ we perform
the depreciation process until this critical condition is
reached. In the simplest case where $\alpha=0$, if
$\gamma>3$, the ratio
$\langle{k^\prime}^2\rangle/\langle{k^\prime}\rangle$
converges to a finite value as $N_v\to\infty$. In this case
$f_c>0$, characterizing a fragile network regime. On the
other hand, when $\gamma<3$, one obtains
$\langle{k^\prime}^2\rangle\propto{k_{max}^{3-\gamma}}$,
where $k_{max}$ is the largest degree of a finite
network. If no other constraint is imposed,
$k_{max}\propto{N_v^{1/(\gamma-1)}}$~\cite{havlin1,moreira},
resulting in $f_c \propto N_v^{-(3-\gamma)/(\gamma-1)}$, for
$2<\gamma<3$. As long as $\gamma<3$, $f_c\to{0}$ as
$N_v\to\infty$, characterizing the robust network regime.

For CB ($\alpha>0$) it is possible that $\gamma\le{3}$ while
${\gamma^\prime>3}$. Since the tail of the distribution at large
values of $k^\prime$ decays as $P^\prime(k^\prime)\propto
{k^\prime}^{-\gamma^\prime}$, the second moment $\langle
{k^\prime}^2 \rangle$ no longer diverges and a robust network
becomes fragile under CB. On the other hand, for PB ($\alpha<0$),
$\gamma^\prime<\gamma$, and the larger exponent, $\gamma$, should
control the decay of the tail of the degree distribution.
Therefore, one may think that for $\gamma>3$, $\langle
{k^\prime}^2 \rangle$ also does not diverge under PB and that a
fragile network cannot turn robust. This simple reasoning is
mistaken, however, as we show in the following.

It is possible that the crossover scale $k_s$ becomes so large
that in practice it does not influence a finite network.  That is
the case, for instance, of the distribution for $f=0.1$ and
$N_v=10^5$, shown in the inset of Fig.~\ref{f.dist}. It may be the
case that, at the critical point of PB failure, a network with
$\gamma^\prime\le{3}$ never displays an observable crossover,
irrespective of the system size, that is,
\begin{equation}
k_s(f_c(N_v))>k_{max}(N_v). \label{eq.assump}
\end{equation}
When Eq.~(\ref{eq.assump}) holds true, Eq.~(\ref{eq.kprime}) may
be rewritten as $k^\prime=C(f_c) k^{1-\alpha}$. In this limit one
can find a linear relation between the parameter $C(f_c)$ and the
occupation fraction
$f_c=\langle{k^\prime}\rangle/\langle{k}\rangle=
C\langle{k^{1-\alpha}}\rangle/\langle{k}\rangle$. The second
moment can be identified as
$\langle{k^\prime}^2\rangle=C^2(\langle{k}^{2-2\alpha}\rangle-\langle{k}^{1-2\alpha}\rangle)+C\langle{k}^{1-\alpha}\rangle$.
From the critical condition
$\langle{k^\prime}^2\rangle/\langle{k^\prime}\rangle>2$ we get
\begin{equation}
f_c=\frac{\langle{k^{1-\alpha}}\rangle^2}
      {\langle{k}\rangle\left[\langle{k^{2-2\alpha}}\rangle-\langle{k}^{1-2\alpha}\rangle\right]}.
\label{eq.crit}
\end{equation}
As long as $2<\gamma^\prime$ and $2<\gamma$, the moments
$\langle{k^{1-\alpha}}\rangle$ and $\langle{k}\rangle$ should both
converge to finite values independent of $N_v$. The moment
$\langle{k}^{1-2\alpha}\rangle$ may or may not diverge, but at
large scales it will grow slower than
$\langle{k^{2-2\alpha}}\rangle\to{k_{max}^{3-2\alpha-\gamma}}$.
Thus, considering that
$k_{max}\propto{N_v^{1/(\gamma-1)}}$~\cite{havlin1}, we arrive at
the behavior
\begin{equation}
f_c\propto N_v^{\frac{\gamma^\prime-3}{\gamma^\prime-1}}.
\label{eq.toma}
\end{equation}
This result shows not only that PB can turn a fragile network
robust but also that the critical exponent with which the
threshold $f_c$ approaches zero is the same as expected for normal
percolation ($\alpha=0$) for a network with a degree distribution
decaying as $P(k) \propto k^{-\gamma^\prime}$.

We can now check the self-consistency of our initial assumption,
Eq.~(\ref{eq.assump}), that networks with $\gamma^\prime\le{3}$ at
the critical point of PB failure do not present a crossover. As
mentioned before, the crossover scale is given by
$k_s\propto{C(f)^{1/\alpha}}$. At the critical point
Eq.~(\ref{eq.toma}) then implies $k_s\propto{N_v^{(\gamma^\prime-3
)/[\alpha(\gamma^\prime-1)]}}$, while
$k_{max}\propto{N_v^{1/(\gamma-1)}}$. From this we obtain
$(k_s/k_{max})^{-\alpha(\gamma-1)}\propto{N_v^{3-\gamma-\alpha}}$.
As long as $\alpha<3-\gamma$, the crossover scale grows faster
than the maximum degree, implying that critical networks with
$\gamma^\prime\le{3}$ and sufficiently strong PB do not display a
crossover in their degree distributions. However, for weak PB,
$3-\gamma<\alpha<0$, Eq.~(\ref{eq.assump}) is violated and the
network may remain fragile.

\begin{figure}[t]
\includegraphics*[width=7.0cm]{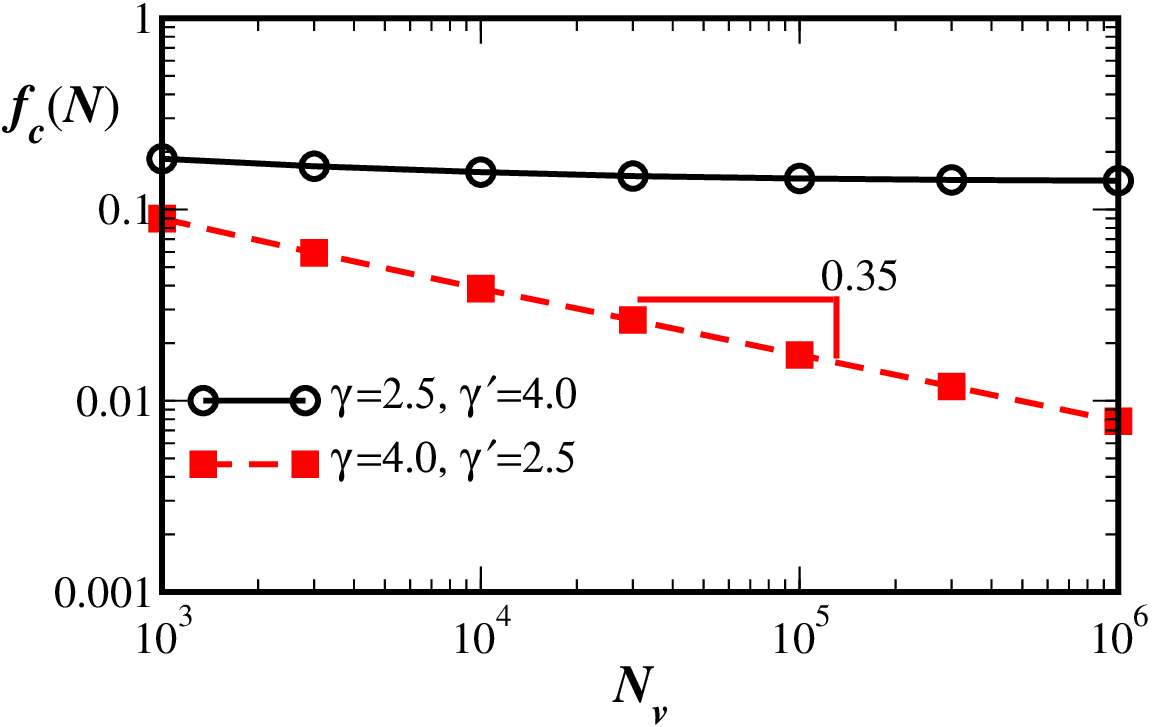}
\caption{The values $f_c$ of the fraction of edges in the
depreciated network at the critical condition as a function of the
network size $N_v$. Criticality is defined as the point where
$\langle{k^\prime}^2\rangle/\langle{k^\prime}\rangle=2$~\cite{havlin1},
To compute this critical fraction we average over $10^4$ network
realizations for each set of parameters. For each network we apply
the percolation processes 100 times.  For networks with
$\gamma=2.5$ submitted to CB with an effective value
$\gamma^\prime=4$ (continuous black line), we observe that the
critical fraction $f_c$ converges to a finite value as $N_v$
grows, confirming the conjecture that a robust network may turn
fragile under CB. The opposite case, a network with $\gamma=4$
submitted to PB with an effective $\gamma^\prime=2.5$ (dashed red
line), has a critical fraction that decays with $N_v$ as a power
law, $f_c\propto{N_v^{-1/\nu}}$. The best fit to the data in this
case results in $1/\nu={0.35\pm{0.02}}$, consistent with the value
$1/3$ expected from Eq.~(\ref{eq.toma}). This result shows that a
fragile network under PB can behave in the same fashion as a
robust network with a degree distribution controlled by
$\gamma^\prime$ under random failure. } \label{f.pc}
\end{figure}

Figure~\ref{f.pc} shows numerical results confirming that a robust
network with $\gamma = 2.5$ submitted to CB failure with $\alpha =
0.5$ turns fragile. In contrast, the second set of results
demonstrates that a fragile network turns robust even for $\alpha
= 3-\gamma = -1$, which is on the borderline between weak and
strong PB.

Our assumption that the probability of failure depends on
degree $k$ can be justified in different contexts. In
artificial networks, e.g., air transportation~\cite{airp},
the capacity of the nodes scales with $k$. Depending on
whether $k$ or capacity grows faster, this system should be
better modelled by CP or PB failure, respectively. Software
systems~\cite{soft} and metabolic networks~\cite{bio}
consist of many agents/nodes acting together in some
function. If all agents are needed, the lack of any of them
can interrupt the process. Alternatively, if any of the
agents can start it, only removal of all edges halts the
process. In both cases, depending on $k$, the edges turn
more fragile or robust. Further, if disrupting the network
is desirable, as in gene fusion networks of cancer
development~\cite{cancer} or terrorist networks, the design
of a dynamical process that targets links between the most
connected nodes (CB) would be more efficient to globally
break down the system. Also, to reduce the risk of epidemic
spreading, it is better to disinfect/immunize connections
between hubs than connections between small
(air-)ports. Note that our analysis does not account for
dynamical correlations in the failure process. It can happen
that removal of a single edge triggers a breakdown, even if
this edge only links to one of the least connected nodes
\cite{Samal}.

We conclude that topologically biased failure can have a dramatic
effect on the percolation properties of scale-free networks. For
{\em central bias} (CB, $0<\alpha<1$), the degree distribution
initially decays with the exponent $\gamma$ up to a certain scale
that depends on the fraction of occupied edges, and then crosses
over to a decay with an exponent $\gamma^{\prime} > \gamma$
defined as in Eq.~(\ref{eq.indek}). For {\em peripheral bias} (PB,
$2-\gamma<\alpha<0$) the crossover is also present but with
$\gamma^{\prime}$ controlling the early decay and the exponent
$\gamma > \gamma^{\prime}$ appearing at large degree. Our results
also demonstrate that a robust network, for which the critical
fraction $f_c$ converges to zero as the network grows, may turn
fragile when subjected to CB ($\alpha >0$). Conversely, a fragile
network, for which the critical point is larger than zero at any
system size, may become robust when subjected to strong PB,
$\alpha<3-\gamma$. Fragility or robustness of a network is thus
not only dependent on the exponent $\gamma$ but can be tuned
quantitatively by the exponent $\alpha$ characterizing the
topological bias.

We thank Hans Hooyberghs for discussions, and FWO-Vlaanderen
Project G.0222.02, CCSS, CNPq, CAPES, FUNCAP, and FINEP for
financial support.

\end{document}